# C-AND: Mixed Writing Scheme for Disturb Reduction in 1T Ferroelectric FET Memory

Mor M. Dahan, Evelyn T. Breyer, Stefan Slesazeck, Thomas Mikolajick, *Senior Member, IEEE*, and Shahar Kvatinsky, *Senior Member, IEEE*

*Abstract*—Ferroelectric field effect transistor (FeFET) memory has shown the potential to meet the requirements of the growing need for fast, dense, low-power, and non-volatile memories. In this paper, we propose a memory architecture named crossed-AND (C-AND), in which each storage cell consists of a single ferroelectric transistor. The write operation is performed using different write schemes and different absolute voltages, to account for the asymmetric switching voltages of the FeFET. It enables writing an entire wordline in two consecutive cycles and prevents current and power through the channel of the transistor. During the read operation, the current and power are mostly sensed at a single selected device in each column. The read scheme additionally enables reading an entire word without read errors, even along long bitlines. Our Simulations demonstrate that, in comparison to the previously proposed AND architecture, the C-AND architecture diminishes read errors, reduces write disturbs, enables the usage of longer bitlines, and saves up to 2.92X in memory cell area.

## I. Introduction

Traditional nonvolatile memories use either a floating gate or a charge-trapping layer to store information as charge. However, the integration of floating gate and charge-trapping memory cells into the state-of-the-art high-k metal gate process is becoming extremely complex due to scaling limitations. Moreover, the high write voltages in the 10V-range generate an unfavorable periphery-to-array ratio, especially for smaller array sizes commonly used in embedded memories. Therefore, alternative random-access memory (RAM) concepts, such as resistive RAM (ReRAM), phase change memory (PCM), magnetic RAM (MRAM) [1] and ferroelectric RAM (FRAM) [2], are gaining traction in the industry.

Recently, a single-transistor (1T) ferroelectric field effect transistor (FeFET) memory cell was presented [3]. The FeFET requires only a single device per cell, uses lower write voltages and is more easily integrated in a CMOS process compared to floating gate and charge-trapping flash cells. Integrating a ferroelectric layer within the gate stack of a regular field effect transistor (FET) enables the transistor to store data in the polarization state of the ferroelectric layer.

FeFET-based designs were presented for different applications such as content-addressable memories (CAMs) [4] [5], field-programmable gate arrays (FPGA) building blocks [6], non-volatile flip-flops [7], and mixed logic and memory operations [8]. Different architectures for memory arrays based on FeFETs have been proposed and demonstrated recently. A 1T ferroelectric NOR-type memory architecture might induce specific requirements on the dimensions of the FeFET to enable a correct operation. For example, the Fe-NOR described in [9] demands sub-10 nm FeFETs, which support only a small number of ferroelectric domains and the use of FinFET devices. Previous works suggested to implement two MOSFETs and a FeFET [10] [11] or a single MOSFET and a single FeFET [10] [12] in a 2T-1T and 1T-1T architecture, which build on an AND-type architecture. An AND array configuration of a single FeFET (1T) memory entails an even denser integration [3] and has been realized in advanced technology nodes for 1T arrays [13] [14]. Yet, the AND array configuration suffers from potential disturbs of surrounding cells during the read and write operations of a specific cell [15]. Aside from memory applications, 1T-FeFET arrays are also integrated in machine-learning applications [16]. For machine-learning approaches, the analog nature of multi-domain ferroelectrics is exploited, by considering the programming or erasing of the transistor as continuous processes [17], which further increases their reading sensitivity. Therefore, finding a FeFET memory architecture that circumvents read errors and write disturbs in the memory array is crucial.

In this paper, we propose C-AND, a novel 1T-FeFET memory array configuration, which diminishes read errors. The proposed writing schemes provide two major improvements: 1) application of different schemes for the write operation to circumvent the effect of asymmetric write voltages, and 2) use of different write voltages for '0' and '1' to prevent write disturb in devices with asymmetric switching voltages. Additionally, we present different techniques to read an entire word in a single cycle and to write an entire word in two cycles. By using a simulation model based on real fabricated devices, we show how the read errors that occurred in the AND array are solved and how the write disturbs are reduced with the C-AND architecture and its read and write schemes.

## II. Background

### A. Ferroelectric Field Effect Transistor

Ferroelectricity is a characteristic of materials that exhibit spontaneous electric polarization [18] (built-in dipoles). These materials usually exhibit two stable saturated polarization states, pointing in opposite directions. To switch between the states, an external electric field must be applied.

A FeFET is a field effect transistor that features a ferroelectric layer instead of (or in addition to) the standard

M. M. Dahan and S. Kvatinsky are with the Andrew and Erna Viterbi Faculty of Electrical and Computer Engineering, Technion–Israel Institute of Technology, Haifa 3200003, Israel (e-mail: mordahan@campus.technion.ac.il; shahar@ee.technion.ac.il).

E. T. Breyer, S. Slesazeck and T. Mikolajick are with NaMLab gGmbH, 01187 Dresden, Germany.

This work was partially supported by the European Research Council under the European Union's Horizon 2020 Research and Innovation Program (grant agreement no. 757259), and by the United States Israel Binational Science Foundation (NSF-BSF, grant agreement no. 2015709), and by the European Fund for Regional Development EFRD, Europe supports Saxony, and by funds released by the delegates of the Saxon parliament.

dielectric layer in the gate stack. Considering the two polarization states of the ferroelectric layer, the FeFET either displays a low threshold voltage (programmed state, logic '1') or a high threshold voltage (erased state, logic '0'). By applying a sufficiently high external voltage across the ferroelectric layer (gate-bulk voltage), the polarization can be reversed, thereby changing the conductivity of the transistor channel. In this manner, the threshold voltage of the transistor can be actively manipulated. Figure 1 exemplarily shows a SPICE simulated $I_{DS}$-$V_{GS}$ sweep of an n-type FeFET, as well as the structure of a FeFET with the two stable states of polarization. Note that voltage drop across the ferroelectric layer in the gate stack determines the ferroelectric polarization states, namely the gate-bulk voltage. During the $I_{DS}$-$V_{GS}$ sweep in Fig. 1, the source and bulk terminals are shortened and grounded so $V_{GS} = V_{GB}$ and it is the voltage drop across the ferroelectric layer.

*B. Multi-Domain FeFET Model*

The utilized behavioral FeFET device model for SPICE simulations is based on the time-dependent Preisach model of hysteresis [19]. The polarization of the ferroelectric layer is assumed to be a superposition of individual, non-interacting dipoles. To shorten the simulation time, the density function of the dipoles is assumed to be Gaussian [20] [21] and the polarization $P$ is therefore simplified to

$$P = k \cdot P_s \cdot \tanh\left(\frac{E_{eff} \pm E_c}{2\delta}\right) + P_{off}, \quad (1)$$

where

$$\delta = E_c \left[\ln\left(\frac{1+P_r/P_s}{1-P_r/P_s}\right)\right]^{-1}. \quad (2)$$

The scaling factor $k$ and the offset parameter $P_{off}$ enable the creation of unsaturated polarization loops, i.e., subloops. The + or – sign in (1) introduces non-volatility by separating the polarization states depending on the sweep direction of the applied electric field. Hence, (1) models the storage of information regarding the history of the ferroelectric and takes into account sub-loops of the ferroelectric layer. Parameters $E_{eff}$, $E_c$, $P_r$ and $P_s$ are the effective electric field, the coercive field, the remanent polarization and the saturation polarization, respectively. When applying an external electric field $E_{ext}$, the time-dependency of the effective electric field [19] [22] is reflected by the delay parameter $\tau_{eff}$ in

$$\frac{dE_{eff}(t)}{dt} = \frac{E_{ext}(t) - E_{eff}(t)}{\tau_{eff}}. \quad (3)$$

The full FeFET model consists of a ferroelectric capacitor, with polarization $P$ emulated by the Preisach model, connected in series to an n-FET (28SLP-based n-FET from GLOBALFOUNDRIES [14]), as shown in Fig. 2. Figure 3 presents the measured and the fitted simulated $I_{DS}$-$V_{GS}$ curves for a FeFET featuring channel dimensions of W=L=500 nm, where W and L are the width and length, respectively. Due to sensitivity limitation of the measurement equipment, the off-current is limited to around 50 nA in the measurements. Therefore, the fitting was done concerning currents above the given limit and the threshold voltages of the high- and low-$V_t$ state. Simulation results were compared to measurements from [14], to ensure real device behavior for the entire current range.

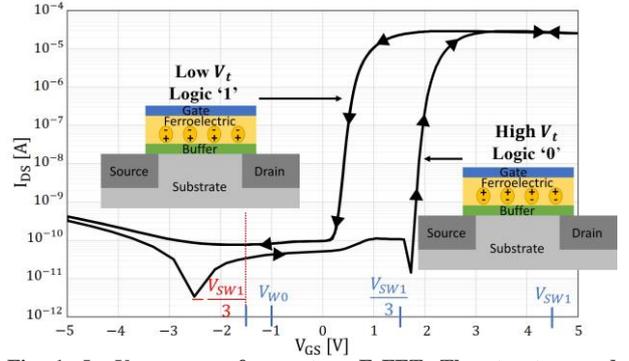

**Fig. 1.** $I_{DS}$-$V_{GS}$ curves of an n-type FeFET. The structure and corresponding polarization direction of each state are shown in the insets. The switching voltage $V_{W0}$ (erase) and applied write voltage $V_{SW1}$ (program) are marked in blue. Note that -$V_{SW1}$/3 is sufficient to write an undesired logical '0'.

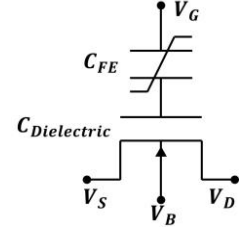

**Fig. 2.** Equivalent circuit of a FeFET model. The FE capacitor and the FET's gate are connected in series.

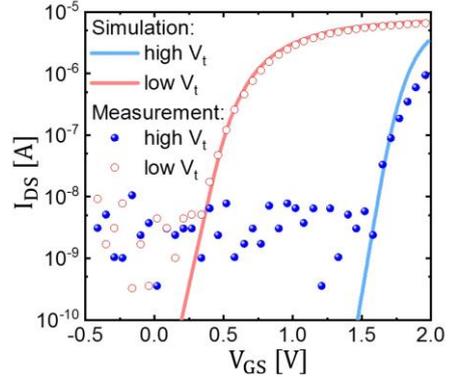

**Fig. 3.** $I_{DS}$-$V_{GS}$ curves of the simulation results compared to the measurement results of the described multi-domain FeFET.

*C. Previously Proposed 1FeFET AND Memory Array*

The AND configuration [3], a special case of a NOR-type memory array, is the most popular design of a single FeFET (1T) memory array. With this configuration, the wordline (WL$_{AND}$) connects the gates of transistors in the same row, while the bitline (BL$_{AND}$) and a separate sourceline (SL$_{AND}$) link the drains and sources of transistors in the same column, respectively. In contrast to the classical NOR array, where the sourceline is grounded, the AND array gives the additional flexibility of driving the SL$_{AND}$ to a specific voltage. Two commonly used writing schemes are the $V_{DD}$/2 and $V_{DD}$/3 schemes, where unselected devices experience only $V_{DD}$/2 or $V_{DD}$/3 to decrease write errors. Usually, only one of the schemes is used for writing both logical states of the FeFET. Commonly, the $V_{DD}$/3 scheme is preferred [3] since the absolute gate-source voltage of all unselected cells will be lower than $V_{DD}$/3 (in contrast to $V_{DD}$/2). Figure 4 shows the AND memory array with the corresponding write and read operations in the $V_{DD}$/3 scheme [3] [23].

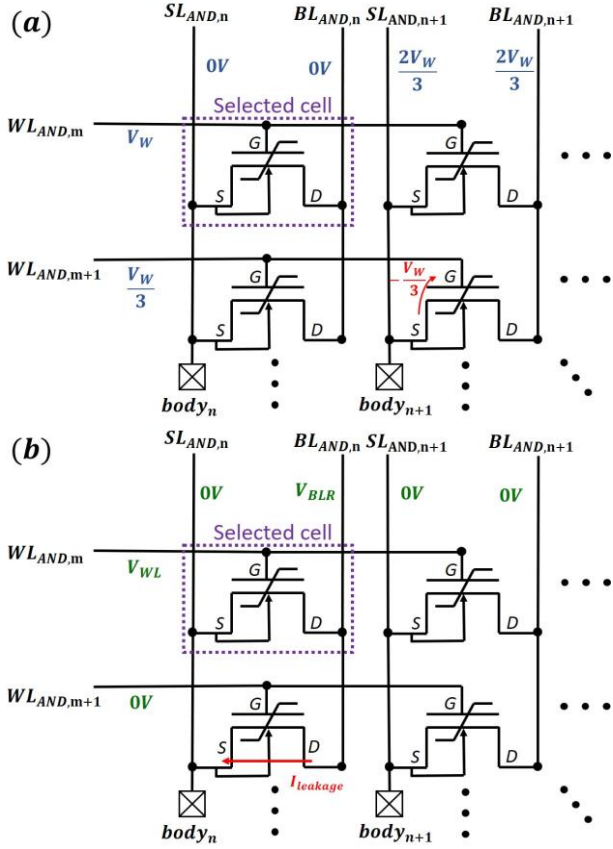

Fig. 4. Structure of the AND array. (a) Write voltages of the $V_{DD}/3$ scheme (blue labels) and (b) read voltages (green labels) are shown. Note that $V_W < 0$ V during erase (write '0' state) and $V_W > 0$ V during programming (write '1' state). The unselected devices in unselected rows and columns (diagonal cells) may suffer from an undesired switching (write disturb).

Yet, the AND architecture suffers from several limitations. During write operations, unselected cells may have a relatively high voltage applied across their ferroelectric layer. Due to possibly asymmetric switching voltages between the polarization states, an unselected cell in an unselected row and unselected column ("diagonal" cell) that experiences $|V_{GS}|=V_W/3$ or $|V_{GS}|=V_W/2$ can be undesirably partially or fully written because the electric field across its ferroelectric layer may be sufficient to write the contrary state [3]. Consider the minimal absolute voltage that causes writing state '0' as $V_{W0}$ and the minimal absolute voltage that causes writing state '1' as $V_{W1}$. Usually, for n-type FeFET, the program voltage is positive ($V_{W1} > 0$ V) while the erase voltage is negative ($V_{W0} < 0$ V), as shown in Fig. 1.

Assume, without loss of generality, that the required absolute voltage to write state '1' is higher than the required absolute voltage to write state '0', namely $|V_{W1}|>|V_{W0}|$. In such a case, writing logical '1' using the $V_{DD}/3$ scheme will bring the diagonal cells to experience a gate-bulk voltage of $-V_{DD}/3$, which may be sufficient to write them to logical '0' (if $|V_{W1}/3|>|V_{W0}|$) or at least will be close to the '0' state switching voltage ($V_{W0}$). This phenomenon can cause undesired full or partial switching to '0' state, as illustrated in Figures 1 and 4(a). Note that this write disturb issue only occurs for the higher absolute write voltage of the two possible states. As we assumed, for $|V_{W1}|>|V_{W0}|$ on third of the voltage needed to write the '1' state ($-V_{W1}/3$) may cause write disturb, while the reversed third voltage required to write the '0' state ($-V_{W0}/3$) is not sufficient to cause write disturb of the unselected cells that stored a logical '0'. With the mixed writing scheme proposed in the next section, this issue is prevented by using different schemes to write the different logical states.

Additionally, like in every NOR-type array, during the read of a selected cell, all the other cells in the same column (same $BL_{AND}$) should not be conductive, otherwise they may cause read errors. Even when applying an inhibitory voltage, such as $V_{GS}=0$ V, to all unselected cells, a low current can leak through the transistors (see Fig. 4(b)) due to a relatively high off current of cells with low-threshold voltages and applied voltage across the transistor channel ($V_{DS}$). Although the leakage current seems to be negligible in smaller arrays, it may cause significant read errors for long $BL_{AND}s$ and $SL_{AND}s$ connecting many cells [15]. The summation of the individual leakage currents at one $BL_{AND}$ or $SL_{AND}$ may lead to reading a logical '1' (high current) rather than logical '0' (low current). This behavior restricts the possible size of the bitlines in the AND arrays, limits the voltages that can be applied and usually demands more complicated sense amplifiers (SA) due to the smaller current reading window. With the proposed C-AND (crossed-AND) architecture and related readout and write schemes presented in the next section, these issues are circumvented, allowing for longer bitlines with a wider current reading window.

### III. C-AND: THE PROPOSED 1T-FeFET MEMORY ARRAY

Ideally, a 1T memory array has different paths for reading and writing. The FeFET device is a four-terminal device and enables the separation of the read and the write paths [24]. Unfortunately, within a 1T memory array, different devices share lines, and hence, there will always be effects from the surrounding cells. We propose a 1FeFET array architecture – denoted "C-AND" - to decrease the dependency between read and write paths. Furthermore, we propose to use the $V_{DD}/2$ and $V_{DD}/3$ schemes in a joint write scheme to address asymmetric switching of FeFET during execution of the write operation, and to reduce the effects from surrounding cells.

#### A. Memory Array

The proposed memory array is shown in Fig. 5. In the proposed architecture, the wordline (WL) connects all the gates in a row and the selectline (SL) connects all the drains of cells in the same row. The bulk line (BuL) connects all the bulk terminals of cells in the same column and the bitline (BL) connects all the source terminals of cells in the same column to a sense amplifier (SA). Note that each column has its own bulk, separated from bulks of other columns. In this structure, the sources and bulks are linked perpendicular to the WL and the drains are linked parallel to the WL. Due to interchanging the position of the drain terminals compared to the AND architecture, the proposed architecture is termed crossed-AND (C-AND). This architecture differs from the NOR architecture that links the drain and bulks perpendicular to the WL and connects the sources parallel to the WL [24].

Interchanging the position of the drain terminals is important, especially for the write operation, since the voltage across the ferroelectric layer (gate-bulk voltage) induces the writing of the specific polarization state. In each device of the C-AND architecture, the writing path is the path between the gate and the bulk of the transistor, and the reading path is situated between the drain and the source of the transistor. Thus, there is full separation between the read and write operations in each memory cell. Additionally, each BL is connected to a sense amplifier (SA) that senses the drain-source current during read operations. With the proposed architecture, there is no current through any device (source-drain current) during the write operation, which prevents extraneous write power. The consumed power results solely from charging the ferroelectric and dielectric capacitors in the FeFET gate stack and from reversing the polarization of the selected FeFET, which makes this architecture a low-power architecture.

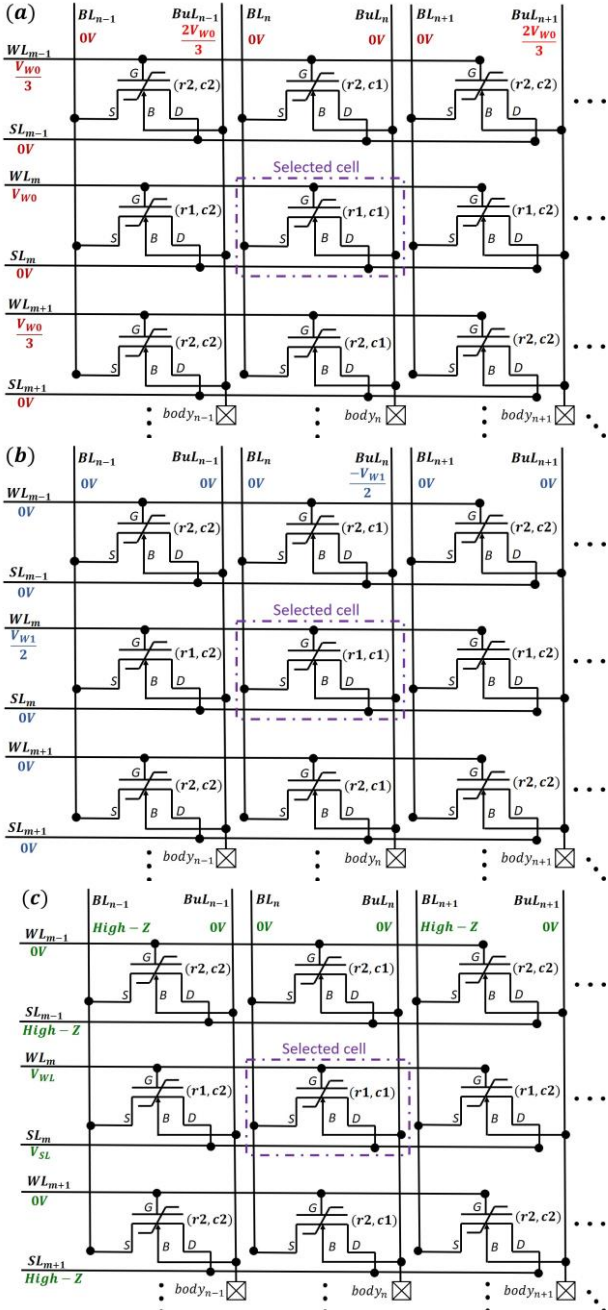

Fig. 5. Schematic of the proposed C-AND memory array with demonstration of write and read schemes: (a) write '0' scheme (red), (b) write '1' scheme (blue), and (c) read scheme (green). (r1,c1) is the selected cell (for read or write), (r1,c2) represents cells in the same row, (r2,c1) denotes cells in the same column and (r2,c2) refers to cells in diagonal, all relative to the selected cell. Note: $V_{W0} < 0$ V and $V_{W1} > 0$ V.

The proposed C-AND memory architecture differs from the previously proposed 1T ferroelectric NOR-type (Fe-NOR) memory array [9] in the following manner: in the C-AND architecture, different columns have different bulks, which grant better selectivity of specific cells, while in the Fe-NOR, all the bulks of all the devices are shortened. Additionally, during write operation in the proposed architecture, there is no need to rely on the drain voltage, while in the Fe-NOR, it is crucial. Furthermore, the Fe-NOR demands sub-10 nm FeFETs which support only a small number of ferroelectric domains.

### B. Write Operation

Table I lists the voltages applied to write logical '0' and '1' states, using the proposed joint $V_{DD}/2 – V_{DD}/3$ write scheme within the C-AND architecture. Figures 5(a) and 5(b) illustrate this write scheme in a memory array with the selected cell and unselected cells (red label in Fig 5a for write '0' operation and blue label in Fig. 5b for write '1' operation).

To avoid write disturb, any unselected cell will ideally experience no voltage across the ferroelectric layer (gate-bulk voltage), i.e., $|V_{GB}| = 0$ V. As this is practically impossible for an array with a single device for each cell and shared lines, we use $V_{DD}/2$ and $V_{DD}/3$ write schemes to ensure that $|V_{GB}|$ is sufficiency low for all unselected cells. The difference between these schemes is that in the $V_{DD}/3$ scheme, unselected cells can experience both $V_{GB} = V_{DD}/3$ and $V_{GB} = -V_{DD}/3$, while in the $V_{DD}/2$ scheme, they may experience $V_{GB} = 0$ V or $V_{GB} = V_{DD}/2$ only (no voltage with inverse polarity). For FeFETs where $|V_{W1}|>|V_{W0}|$, only the writing of logical '1' can cause write disturb, therefore the scheme utilized to write logical '1' should avoid applying voltages of inversed polarity. Hence, state '1' with write voltage $V_{W1}$ is written with the $V_{DD}/2$ scheme to ensure that there are no voltages with inverse polarity, while the state '0' with write voltage $V_{W0}$ is written with the $V_{DD}/3$ scheme, since the voltage of inverse polarity is insufficient to cause write disturb.

For example, assume FeFET switching voltages are $V_{W1} = 4.5$ V and $V_{W0} = -1$ V (asymmetric switching voltages). If we use the $V_{DD}/3$ scheme to write state '1', diagonal cells in the array will experience $V_{GB} = -V_{W1}/3 = -1.5$ V, which entails an unwanted overwriting of the stored data in these cells (see Figures 1 and 4(a)). Even if the voltage to write the FeFET to state '1' was much lower (for instance $V_{W1} = 2.1$ V), the diagonal cells would experience an absolute voltage higher than $V_{DD}/2$ ($V_{GB} = -0.7$ V). This value is approximately the voltage required to set the FeFET into the '0' state. Accordingly, it can induce a partial polarization reversal in the ferroelectric layer of the FeFET, which finally writes a logical '0' into unselected cells by switching a significant number of domains, such that the stored data would be overwritten. The proposed joint write scheme mitigates the overwriting of unselected cells independent of the switching voltages of the FeFETs, provided the write voltages, $V_{W0}$ and $V_{W1}$, are properly chosen.

The voltages $V_{W0}$ and $V_{W1}$ should be chosen as minimal absolute voltages that are sufficiently high to change the polarization direction of the ferroelectric layer concerning a reasonable delay. Additionally, it has to be ensured that $V_{W0}/3$ and $V_{W1}/2$ are insufficient to undesirably change the polarization state of the FeFET. It is possible to use higher absolute write voltages than necessary to expand the read window, however, the power consumption would increase. Additionally, applying a higher absolute write voltage may impair the endurance and may finally lead to a breakdown of the device. Note that with the proposed writing scheme, the drain-bulk and the source-bulk diodes are always in zero or reversed bias, preventing unwanted significant bulk currents.

TABLE I.
WRITE SCHEME IN C-AND ARCHITECTURE
($V_{W0} < 0$ V, $V_{W1} > 0$ V)

| Operation | Write logic '0' | | Write logic '1' | |
|---|---|---|---|---|
| Row Line | Selected Row | Unselected Row | Selected Row | Unselected Row |
| WL | $V_{W0}$ | $V_{W0}/3$ | $V_{W1}/2$ | 0 |
| SL | 0 | 0 | 0 | 0 |
| Col Line | Selected Columns | Unselected Columns | Selected Columns | Unselected Columns |
| BuL | 0 | $2V_{W0}/3$ | $-V_{W1}/2$ | 0 |
| BL | 0 | 0 | 0 | 0 |

TABLE II.
READ SCHEME IN C-AND ARCHITECTURE

| Row Line | Selected Row | Unselected Row |
|---|---|---|
| WL | $V_{WL}$ | 0 |
| SL | $V_{SL}$ | High-Z |
| Col Line | Selected bits | Unselected bits |
| BuL | 0 | 0 |
| BL | 0 | High-Z |

Furthermore, it is possible to program or erase multiple cells in the same row by applying 0 V to all corresponding BLs for write '0' operation or by applying $-V_{W1}/2$ to all corresponding BuLs for write '1' operation (Fig. 5(a) and 5(b), with more than one selected cell in the same row). In the first cycle, state '0' is written in all selected cells, while in the following cycle, state '1' is written into the remaining cells of the same row, resulting in writing of an entire word in two cycles, regardless of the word size (the order of writing '0's and writing '1's can be switched without affecting the number of needed cycles).

*C. Read Operation*

To read bits along a certain wordline, a read voltage $V_{WL}$ is applied to the WL connecting cells in a specific row that stores the word to be read. This is achieved by setting the WL voltage to be between the threshold voltages of the '0' and '1' states ($V_{t0}$ and $V_{t1}$, respectively), i.e., $V_{t0}<V_{WL}<V_{t1}$. Table II lists the voltages applied to the word, select, bulk and bit lines. Due to the selected drain voltage $V_{SL}$, the FeFET is read out in the saturation regime of the transistor. Figure 5(c) illustrates this read scheme in the C-AND memory array (green label) for reading the bit of the memory cell in the middle of the drawn nine-cell array segment. To determine the logical value of the readout current, the SA acts in a current-mode sensing scheme where the sensed current is compared to a reference current source [25]. The reference current is chosen to be between the low current (corresponds to $V_{t0}$) and high current (corresponds to $V_{t1}$) so the sensed current is labeled as logical '0' and '1' according to the polarization state and the threshold voltage of the sensed device. Usually, with NOR-type non-volatile memories (NVM), a single byte or word is read in a single cycle. C-AND also enables the read of several bits along the same wordline by grounding the BLs of all desired cells in the selected row. Since each BL is connected to a SA, all the selected cells in the selected row will be read simultaneously within a single cycle.

The proposed read scheme solves the read errors that might occur in AND arrays. On each selected BL, the only device that experiences $V_{SL}$ across its channel and gate-bulk voltage of $V_{WL}$ is the selected cell. Since the current along BL is not summed up (unlike in the AND architecture), the memory array can contain more rows than the AND architecture without being limited by reading disturbs. The physical limitation on the number of rows is primarily due to the voltage drop across the BLs (non-ideal wires), not due to an architecture limitation.

*D. Physical Design and Area*

The physical designs of the AND and the C-AND arrays are shown in Fig. 6. For the C-AND array, poly-silicon wires and metal1 (M1) lines connect the gates and the drains of each row, respectively, while metal2 (M2) links the sources of each column. Each column has its own bulk (p-well), which is shared with all the devices in the same column. The density of the array can be increased by isolating the well by a deep trench isolation together with a buried BL, as discussed previously for flash devices [26]. The C-AND array can be implemented using a single metal layer by connecting each row's gates and drains using poly-silicon wires and diffusion (active) lines, while M1 links the sources of each column. In this manner, the resistance of the SLs (connected by diffusion) is higher compared to the resistance obtained when using an additional metal layer.

To evaluate the effective cell size of the C-AND architecture and to compare it to the cell area in the AND architecture, we designed the layout of the two architectures in the GLOBALFOUNDRIES 28SLP (where $\lambda = 28$ nm) using Virtuoso Layout Suite GXL tool, as shown in Fig. 6. In contrast to a logic process, where $\lambda$ defines the technology gate dimension, in a memory process, $\lambda$ defines the minimum half-pitch of M1, which in the discussed technology is approximately 50 nm. Therefore, comparing the minimum cell size to what could be achieved in a logic process would be unreasonable because this comparison would limit the cell size to much larger values than expected for a memory process. As a point of reference, the minimum feasible cell area in a memory process is $4\lambda^2$. When using $\lambda$ as the technology node name of a logic process, the minimum cell area would be in the order of $16\lambda^2$ (assuming both directions are limited by the minimum metal pitch). The cell area in the AND array is $244.14\lambda^2$, while in C-AND, it is only $83.57\lambda^2$ – an improvement of 2.92X. The spacing between the different bulks is important and stands at $35.7\lambda$ in this technology (triple-well) but can be reduced to $9\lambda$, as reported in [27], to further reduce the overall array area. Table III presents the comparison of the area with and without the spacing between the different bulk.

*E. Leakage Paths*

In the proposed C-AND architecture, there may be leakage currents from the selected SL to the selected BLs through three series unselected devices. These leakage currents start from the selected SL, go through unselected devices in the selected row (which in the worst case are open and have low resistance) to unselected BLs. From the unselected BLs the currents go through closed devices to unselected SL (from the source to the drain) and then through another close device to the selected BL. Figure 7 presents two of the leakage paths. The resistance ratio between the devices in the leakage path and the device in the desired path determines whether these leakage currents, which sum up on the selected BL, can cause read error. The worst-case scenario is that the selected device is on high resistance $R_S$ (stored data is logical '0'), and the unselected devices are storing logical '1'. In that case, the devices in the unselected row are ideally non-conductive ($V_{GB} = 0$ V) and have a resistance of $R_C$, and the unselected device in the selected row is conductive ($V_{GB} = V_{WL}$) and has a resistance of $R_O$. The channel resistance in the desired path is $R_S$, while the effective resistance ($R_{eff}$) of all leakage paths for an array with $m$ rows and $n$ columns is

$$R_{eff} = \frac{R_O}{n-1} + \frac{R_C}{(m-1)\cdot(n-1)} + \frac{R_C}{m-1}. \quad (4)$$

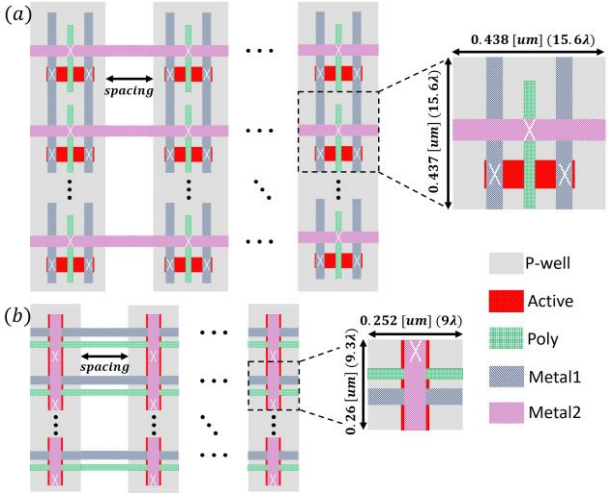

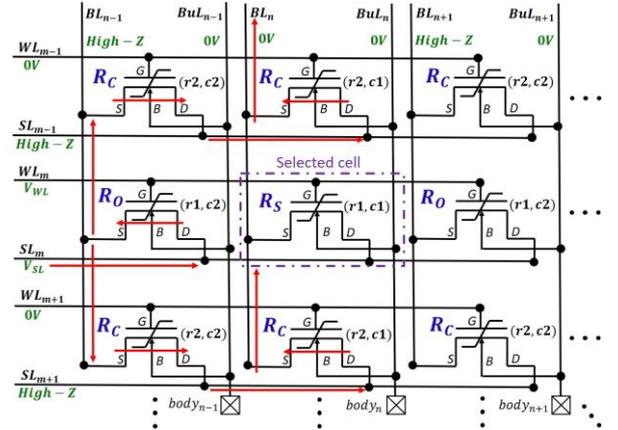

Fig. 6. Layout and cell structure of the (a) AND and (b) C-AND arrays. The total area depends on the spacing between different wells. Spacing in 28-SLP GF design rules stands at $35.7\lambda$ and may be substantially smaller in other technologies. $\lambda$ is defined according to the logic process naming, here 28 nm. The cell area of the AND and C-AND is, respectively, $244.14\lambda^2$ and $83.57\lambda^2$.

Fig. 7. Leakage paths during read operation. Leakage current starts from unselected device in the selected row (device (r1,c2)), go through unselected device in unselected row and column (device (r2,c2)) and then to unselected device in selected column (device (r2,c1)). Two of these paths are presented in with red arrows.

TABLE III.
COMPARING CELL AREA WITH AND WITHOUT SPACING

| Spacing | AND | C-AND | Improvement |
|---|---|---|---|
| With | $801.54\lambda^2$ | $415.2\lambda^2$ | 1.93X |
| Without | $244.14\lambda^2$ | $83.57\lambda^2$ | 2.92X |

The last term in (4) is the largest term since the resistance $R_O$ is a resistance of a conductive transistor while $R_C$ is a resistance of a non-conductive device. Therefore, an upper bound to the effective resistance would be

$$R_{eff} > \frac{R_C}{m-1}. \quad (5)$$

Hence, the undesired current from leakage paths is bounded linearly with the number of rows and depends on the channel resistivity of unselected closed device. Ideally, the resistance of closed device is infinitely large, and no leakage current is donated. Practically, there will always be leakage currents, so the memory size that can be supported depends on the SA sensitivity and the closed channel resistance of the specific technology. Still, the possible BL size of the C-AND architecture will be larger than the possible BL size of the AND architecture since for the same voltage drop, in the AND architecture the summed-up currents are currents through a single device while in the C-AND are currents through three devices.

*F. Power Analysis*

During the writing operation, all SLs and BLs are grounded and hence, all the transistors in the array experience $V_{DS} = 0$ V, resulting in the absence of current through the device's channels during the write operation and hence, to no extraneous power consumption. The only current (and power) consumed is that for the charging and discharging of the gate capacitors, which include the ferroelectric capacitance and the dielectric capacitance (see Fig. 2). Hence, the overall power consumed during the write operation is used for the operation of storing the data in the polarization states of the FeFETs in a nonvolatile manner. The power consumption depends on the area of the FeFET devices (length and width of the FeFET channel), on the resistivity and capacitance of the poly-silicon wires (which depends on the spacing), and on the write voltage ($V_{W0}$ or $V_{W1}$).

For the read operation, the WLs of unselected rows are grounded. Hence, these WLs are not charged and, correspondingly, no power is consumed. With the proposed read scheme, the selected SL is set to $V_{SL}$ and the selected BLs are grounded while all other SLs and BLs are floating. The devices in the selected row experience $V_{DS} = V_{SL}$ and the current through the selected SL ($I_{DS}$) depends on the stored word. All the transistors in the same row are connected in parallel during the read operation (same $V_{DS}$ and $V_{GB}$) and the current through each transistor channel depends on the stored value. For a word with $n_0$ bits corresponding to '0' (with off-current of $I_{low}$) and $n_1$ bits corresponding to '1' (with on-current of $I_{high}$), the current through the selected SL ($I_{SL}$) is

$$I_{SL} = n_0 \cdot I_{low} + n_1 \cdot I_{high}, \quad (6)$$

and the power consumed by the selected devices ($P_{SL}$) is

$$P_{SL} = I_{SL} \cdot V_{SL}. \quad (7)$$

The highest current (and power) is obtained in cases where the word contains only bits of logical '1' state. In this case, the total power of an $n$-bits word ($P_{SL,max}$) is

$$P_{SL,max} = I_{SL} \cdot V_{SL} = n \cdot I_{high} \cdot V_{SL}. \quad (8)$$

In addition to the power resulting from the currents through the channels of the selected transistors, the power consumption related to charging the selected WL to the read voltage $V_{WL}$ (namely $P_{WL}$) depends on the WL resistance and capacitance as well as the gate stack capacitance of the transistors. Additionally, the consumed power from leakage paths and leakage to the bulks, $P_{leak}$, should also be added. In conclusion, the total power of a read operation consumed for read of a single device is

$$P_{read\_bit} = P_{SL} + P_{WL} + P_{leak}. \quad (9)$$

For read operation of *n* bits of logical '1', the consumed power would be

$$P_{read\_bit} = n \cdot P_{SL} + P_{WL} + P_{leak}. \quad (10)$$

The power analysis reveals that during write, the C-AND architecture consumes power only for data storing. During the read operation, the consumed power is due to the desired read operation and undesired leakage currents. The power consumption as a result of leakage paths can be reduced by improved technology with higher off resistance.

## IV. METHODOLOGY AND EVALUATION

### A. Simulation Methodology

To evaluate the C-AND architecture, we simulated the memory array with the corresponding read and write schemes, using the FeFET model described in Section II-B, with model parameters calibrated to the transfer curves of manufactured devices, as listed in Table IV. The line parasitics were extracted from the physical layout. The ferroelectric capacitor in the model was implemented in Verilog-A and the transistor model is the 28SLP-based n-FET from the GLOBALFOUNDRIES library [14]. Cadence Virtuoso was used to conduct the simulations.

### B. Long Bitlines

To evaluate the effect of long BL during write operation, we simulated a single column of different sizes, from two to 2048 rows. It is sufficient to model only a single column since the current through each BL in the AND architecture is the summation of the current through all cells in the same column. For the C-AND architecture, the leakage current from leakage paths was added, assuming the memory array has the same number of rows and columns. The current through the FeFET during read operation depends only on the polarization state stored in each FeFET-based memory cell. The worst-case scenario is the readout of state '0' (low current) from a specific cell, while all unselected cells carried '1's (namely, low threshold voltage). The leakage of these cells sums up, and cannot be neglected for long BLs.

Figure 8 compares the readout currents of the AND and C-AND architectures for a varied number of rows. The difference between the readout of logical '1' and logical '0' is the read window for each architecture. The simulations were performed for two to 2048 cells in a row. A trendline continues the general tendencies for longer columns. As expected for the AND array, for the worst-case scenario, leakage currents of the unselected cells summed up to approximately 30 nA for a column size of 2048 cells. This represents a considerable read error when compared to 400 nA in case of a regular readout of the '1' state. For C-AND, the current readout of the '0' state of a selected cell on BL with 2048 cells was only 40 pA, since the C-AND and AND has the same voltage drop $V_{SL}$, and the leakage currents in the C-AND pass through three transistor channels in series (which at least two of them are non-conductive), while for the AND the leakage current path is only through a single non-conductive transistor channel. Thus, readout errors were eliminated and an $I_{on}/I_{off}$ ratio of 10$^4$ is achieved for a column size of 2048 cells. For the C-AND architecture, the $I_{on}/I_{off}$ ratio is decreasing much slower than the AND architecture, maintaining the high $I_{on}/I_{off}$ ratio for different column sizes.

TABLE IV.
MODEL PARAMETERS AND PARASITIC

| Parameter | Description | Value | units |
|---|---|---|---|
| $L$ | Transistor channel length | 500 | nm |
| $W$ | Transistor channel width | 500 | nm |
| $T$ | Ferroelectric thickness | 10 | nm |
| $P_s$ | Saturation polarization | 0.2 | C/m$^2$ |
| $P_r$ | Remanent polarization | 0.19 | C/m$^2$ |
| $V_c$ | Coercive voltage | 1.04 | V |
| $R_m$ | Metal resistance | 9.45 | Ω/μm |
| $C_m$ | Metal capacitance | 0.22 | fF/μm |
| $R_p$ | Polysilicon resistance | 2000 | Ω/μm |
| $C_p$ | Polysilicon capacitance | 0.15 | fF/μm |
| $V_{W0}$ | Write '0' voltage | -1.5 | V |
| $V_{W1}$ | Write '1' voltage | 3.2 | V |
| $V_{WL}$ | WL voltage for read | 1 | V |
| $V_{SL}$ | SL voltage for read | 1 | V |
| $t$ | Duration of read and write | 10 | μs |

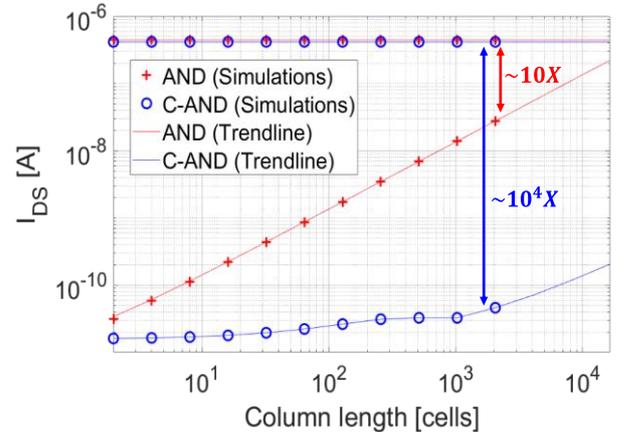

Fig. 8. Comparison of an AND array (red) to the C-AND architecture (blue) for different column sizes. Read window in AND architecture is decreasing faster than the C-AND read window.

### C. Write Disturb in C-AND

As illustrated in Fig. 5, in any operation, there are four groups of cells defined by their applied voltage: *(r1,c1)*, *(r1,c2)*, *(r2,c1)* and *(r2,c2)*. Each cell in these groups can be in one of two different states: state '0' or state '1', so there are 16 different possible cases (each cell in the four groups can be in either state '0' or state '1' and the write operation can be either write state '0' or state '1'). To examine the effect of a write disturb when using the mixed writing scheme, a 16 by 16 array was simulated. Additionally, the effect of partial voltages applied to all the different cells and states was tested. For the worst-case scenario, we measured the four cells in the corners, which represent the four different groups, as shown in Fig. 9. To test the effect of consecutive writes, the write time was set to 10 μs, where the devices have reached a steady state.

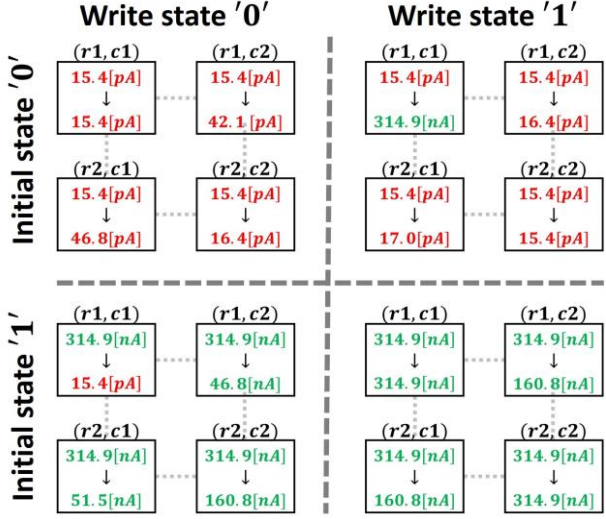

**Fig. 9.** Write operation of a selected cell in a 16 x 16 C-AND array and the influence onto all surrounding cells. The selected cell is labeled by (r1,c1), while unselected cells are labeled by (r1,c2), (r2,c1) and (r2,c2). Different initial states and different write values are presented for all four different cell locations. Currents through cells storing a logic '0' state are indicated in red, while currents through cells storing a logic '1' are indicated in green. The read current of each cell is shown before and after the write operation.

Figure 9 shows the operation of writing '0' and '1', using the proposed joint $V_{DD}/2$-$V_{DD}/3$ scheme, and examines the resulting output current amplitude of cells in all four groups ((r1,c1), (r1,c2), (r2,c1) and (r2,c2)), with all possible previous logical states (state '0' and state '1') and all possible write operations (write logical '0' and write logical '1'). The readout currents before and after the write operation are shown in the squares representing the individual memory cells. Only the stored data in the selected cell was changed when writing a device to the opposite state, while the other cells maintained their state. It can be observed that for writing '1' the current in the diagonal cells did not change. After writing logical '0', however, the current in diagonal cells was increased a bit if the former state was logic '0', while it was decreased to half if the former state was logic '1', but the transistor's logic state remained unchanged. In all 16 different cell locations, with different initial states and different write operations, there was still a separation of three orders of magnitude between the readout current of state '0' to the readout current of state '1'. The highest current representing logical '0' state was 46.8 pA, while the lowest current representing logical '1' state was 46.8 nA. Therefore, the minimal $I_{on}/I_{off}$ ratio was approximately $10^3$.

### D. Writing a Word in Two Cycles

Arrays with eight rows and eight columns were utilized to simulate the programming and erase operation of an entire 8-bit word in two cycles. Figure 10 illustrates how an entire word can be written in two cycles. In the first cycle, state '0' is written to bits 4 to 7, according to the $V_{DD}/3$ scheme (Fig. 10(b) and Fig. 10(c)), while in the subsequent cycle, state '1' is written to bits 0 to 3 in the $V_{DD}/2$ scheme (Fig. 10(b) and Fig. 10(c)). The readout of an entire word in a single cycle is shown in Fig. 10(a) by reading each bit from different BLs. The effect of each write operation on the surrounding cells can also be can be observed in Fig. 10(a). That is, small differences in readout current resulted from the different threshold voltages of the FeFETs (i.e., due to different stored polarization) and from potential write disturbs that caused varying currents for cells that store the same logical state.

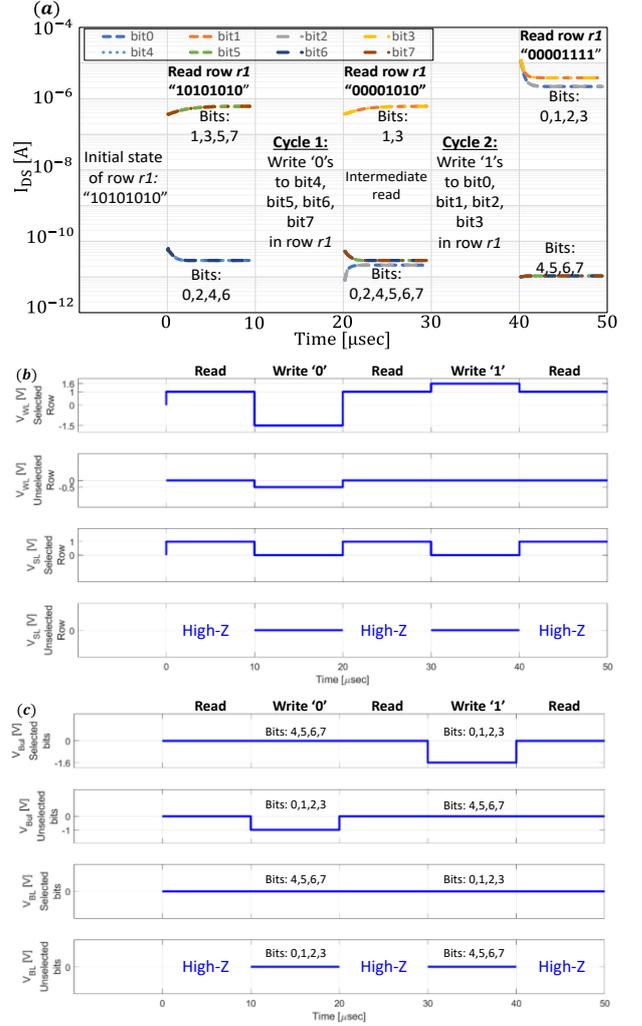

**Fig. 10.** Writing different values to an 8-bit word in an 8 x 8 C-AND array in two cycles. (a) Readout currents during the read operations. (b) Voltage applied to the WLs and SLs of the selected row and unselected rows during the read and write operations. (c) Voltages applied to the BLs and BuLs of the selected and unselected bits (columns) during the read and write operations.

### E. Process Variation

To evaluate the effect of process variation and write voltage variation, we performed Monte-Carlo simulations to randomly choose different conditions on the transistor dimensions. Two rows and two columns arrays, which represent devices from all four groups ((r1,c1), (r1,c2), (r2,c1) and (r2,c2)), were chosen to determine the effect of process variation on any location of a cell relative to the selected cell. To consider the effect of leakage paths, we added the expected leakage current of a 512 by 512 array to the results. The simulation results are based on 9000 tests with standard deviations of 75 mV, 160 mV and 50 nm for the write logical '0' voltage, write logical '1' voltage and the transistor dimensions (width and length of the channel), respectively. The rest of the model parameters are as presented in Table IV.

Figure 11 shows the histograms of the Monte-Carlo simulation. Different write voltages and different dimensions entailed a deviation of the threshold voltage of each device (Fig. 11(a)), and hence, the readout current varied (Fig. 11(b)). Irrespective of the induced variations, there was still an $I_{on}/I_{off}$ ratio of 80, which enables a full separation between the logical '0' and '1' states. When analyzing Fig. 1, Fig. 9, and Fig. 11, it becomes apparent that the $I_{on}/I_{off}$ ratio of a single cell shrinks from approximately $10^5$ (Fig. 1) to approximately $10^3$ when the cell is used in an array arrangement, in which different cells share lines (Fig. 9), and shrinks even further to approximately 80 when taking into account cell variations (Fig. 11). This observation demonstrates the feasibility of the C-AND architecture and poses restrictions on the desired sensitivity of the SA located on each BL.

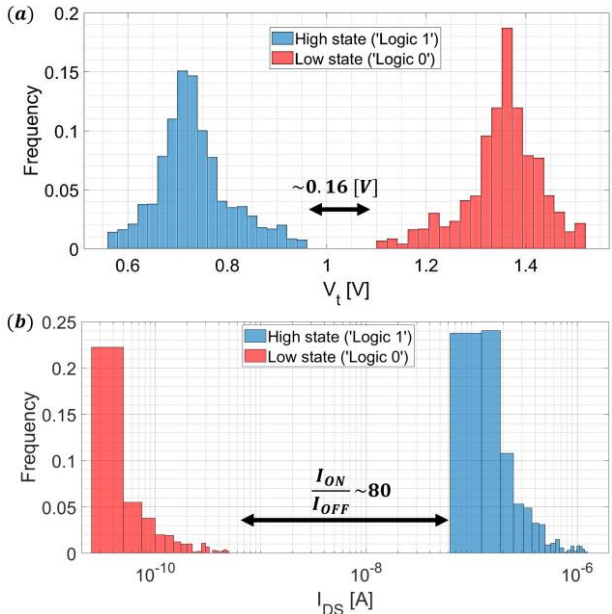

Fig. 11. Monte-Carlo simulations of process variation in the C-AND architecture. There is a full separation (approximately 80X) between the readout current of the '0' and '1' logical states. (a) Threshold voltage variation. (b) Readout current variation.

*F. Power Consumption*

The peak power consumption of the read operation was evaluated for different array sizes (we assume the number of rows and columns is identical), ranging from two to 32. The highest power consumption is obtained for arrays where all cells are in the logical '1' state. Figure 12 shows the peak power consumption of a read operation of a single bit. The leakage power is negligible compared to the read power of the selected cell. Our results show that the read power of a single bit is almost independent of the memory size.

V. DISCUSSION

*A. Partial Switching*

FeFETs can show an analog behavior [16], i.e., partial switching may occur even at voltages below the coercive voltage (see Figures 9 and 10). This behavior depends on the number of domains within the ferroelectric layer. In the following, the individual domain sizes are assumed to be 500 $nm^2$. A device with W/L=50 nm/50 nm would accommodate 5 domains, i.e., the switching will exhibit an abrupt, rather than analog, behavior, and therefore, partial switching is negligible. However, for a device with W/L= 500 nm/500 nm comprising approximately 500 domains, the device will show a gradual switching, i.e., analog behavior [17].

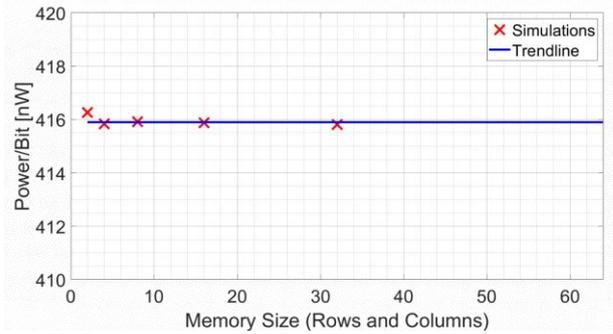

Fig. 12. Peak power for a single bit read versus the array size. Each array has the same number of rows and columns.

The phenomenon of partial switching is even more critical when considering the trade-off between write voltage and write time [28]. Longer write times lower the voltage at which a certain domain will change its polarization state. The partial switching issue becomes dominant when cells are partially selected ($V_{DD}/3$ or $V_{DD}/2$) for numerous cycles. As a result, the cells are gradually overwritten. Thus, each time an absolute disturb voltage of $V_{DD}/3$ or $V_{DD}/2$ is routed to the gate of an unselected cell, the corresponding FeFET device is slightly programmed or erased, and after a sufficient number of cycles, these effects may accumulate and change the state of the FeFET [29]. Applying active compensation mechanisms (such as a refresh operation) to the memory can solve this problem of non-uniform access patterns to read and write cells, at the cost of higher power consumption and lower access rate to the memory. Furthermore, splitting the memory into smaller arrays can help as fewer cells would suffer from write disturb. It can be observed from Fig. 8 that even for the C-AND architecture, the bigger the array size is, the smaller the readout current window becomes.

*B. Scaling of the FeFET*

In non-volatile memories (NVM), the source and the drain of the transistor experience voltages which are higher than the standard process voltage. This issue limits the scaling of the memory devices. The write operation with the C-AND architecture does not require hot carriers or current through the transistor's channel and it only depends on the ability to control the voltage across the ferroelectric layer, *i.e.*, the gate-bulk voltage. However, scaling of the channel of the transistor can cause a punch-through [30], *i.e.*, the effective body voltage will be a function of the source and drain voltages. In such a case, the control of the channel potential via the bulk contact, *i.e.*, relying on bulk biasing, will be ineffective. This imposes restrictions on the scaling and the geometry of the FeFET devices suitable for implementing the C-AND architecture.

VI. CONCLUSION

$HfO_2$-based ferroelectric field-effect transistors exhibit several desirable features, such as CMOS compatibility, fast switching, good scalability, low-power, and non-volatility. In this paper, we presented the C-AND architecture, which exploits the unique properties of the FeFET to design a novel memory array structure based on a single FeFET in each memory cell. We propose a write operation scheme that addresses the potentially asymmetric switching voltages of the FeFET by combining the $V_{DD}/3$ and $V_{DD}/2$ write schemes to utilize different absolute write voltages. The C-AND architecture enables the writing of an entire word in two consecutive cycles. The depicted structure is suitable for manufacturing, with the potential for lowering the cell size by using self-alignment techniques for the diffusion lines and the wells. Power analysis indicates a low power memory architecture with minimal leakage.

A comparison of the C-AND array to the AND array indicated that the read error, caused by leakage currents, can be prevented even when using long BLs, and that the cell area can be lowered by up to 2.92X. Reliable memory architecture is crucial to realize the potential of FeFET-based memories. The C-AND architecture has the potential to be the platform for future non-volatile FeFET-based memories and a candidate to replace the floating gate and charge-trapping memories.

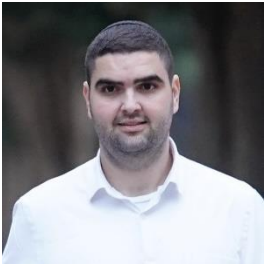**Mor M. Dahan** is a Ph.D. candidate at the Andrew and Erna Viterbi Faculty of Electrical and Computer Engineering, Technion – Israel Institute of Technology. Mor received his B.Sc. and M.Sc. degrees in Electrical and Computer Engineering in 2018 and 2021, respectively, both from the Technion – Israel Institute of Technology.

His research is focused on utilizing ferroelectric devices for both non-volatile memory and logic. Mor's main research interests include advanced electrical characterizations of ferroelectric field-effect transistors (FeFETs), developing FeFET-based in-memory computing capabilities, and exploiting FeFETs for neural networks (NN) and artificial intelligence (AI) acceleration.

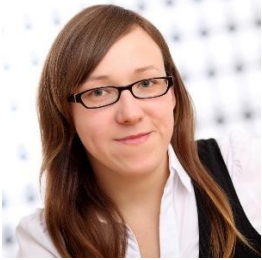**Evelyn T. Breyer** received the Dipl.-Phys. from the Chemnitz University of Technology in 2011. From 2011 until 2013 she joined the Group of Semiconductor Physics at the Chemnitz University of Technology and investigated organic/inorganic heterojunctions. In cooperation with General Electric in Berlin and as part of an integrated degree program she received the B. Eng. degree in applied electrical engineering from the Berlin School of Economics and Law in 2016. Since 2016 she has been a research associate and Ph.D. candidate at the Nanoelectronic Materials Laboratory (NaMLab gGmbH).

Her main research interests include logic-in-memory and nonvolatile logic circuits based on ferroelectric field-effect transistors.

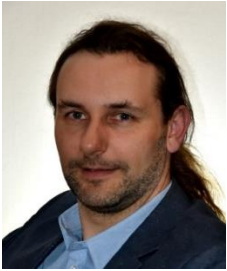**Stefan Slesazeck** received the Ph.D. degree from TU Dresden in 2004. Since 2009 he is a Senior Scientist with NaMLab responsible for concept evaluation, hardware development, electrical characterization, and modeling of memories. On these topics, he is (co)-author of >200 publications and holds 6 US patents.

His research interests comprise the development of novel memory devices with a focus on ferroelectric devices such as FeFETs and FTJs as well as research on novel computing paradigms based on these devices.

Prior to joining NaMLab Stefan was a project leader for the predevelopment of new memory concepts with Qimonda, Dresden, Germany, focusing on concept evaluation for 1T-DRAM, including floating body devices, cell concepts, and access schemes for WL-driver and sense amplifier. As a Device Engineer with Infineon Technologies, he focused on the module development of 3D DRAM access devices in 65-nm and 46-nm buried word line technology and predevelopment of 3D DRAM access devices as FinFET and EUD.

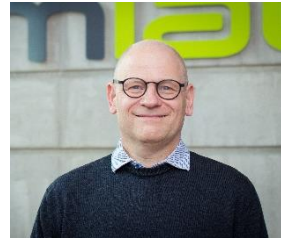**Thomas Mikolajick** (Senior Member, IEEE) Thomas Mikolajick received the Dipl.-Ing. and the Dr.-Ing. In electrical engineering in 1990 and 1996 both from the University Erlangen-Nuremberg.

From 1996 till 2006 he was in semiconductor industry (Siemens Semiconductor, Infineon, Qimonda) developing CMOS processes and memory devices. In that phase he was involved in the development of SBT based ferroelectric memories, the setup of the CBRAM project and the setup of the Infineon standalone Flash activity. In 2006 he was appointed professor for material science of electron devices at TUBAF Freiberg. Since 2009 he is a professor for nanoelectronics at TU Dresden and in parallel the scientific director of NaMLab GmbH. He is author or co-author of more than 450 publications (current h-index of 68 according to google scholar) and inventor or co-inventor in more than 50 patent families.

He is a one of the speakers of the center for advancing electronics Dresden (cfaed). Together with his co-workers at NaMLab and other partners he pioneered hafnium oxide based fluoride structure ferroelectrics which is currently considered to be an important ingredient for future low power nonvolatile memories.

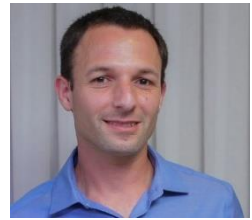**Shahar Kvatinsky** (Senior Member, IEEE) is an Associate Professor at the Andrew and Erna Viterbi Faculty of Electrical and Computer Engineering, Technion – Israel Institute of Technology. Shahar received the B.Sc. degree in Computer Engineering and Applied Physics and an MBA degree in 2009 and 2010, respectively, both from the Hebrew University of Jerusalem, and the Ph.D. degree in Electrical Engineering from the Technion – Israel Institute of Technology in 2014.

From 2006 to 2009, he worked as a circuit designer at Intel. From 2014 and 2015, he was a post-doctoral research fellow at Stanford University.

Kvatinsky is a member of the Israel Young Academy. He is the head of the Architecture and Circuits Research Center at the Technion, chair of the IEEE Circuits and Systems in Israel, and an editor of Microelectronics Journal and Array. Kvatinsky has been the recipient of numerous awards: 2021 Norman Seiden Prize for Academic Excellence, 2020 MDPI Electronics Young Investigator Award, 2019 Wolf Foundation's Krill Prize for Excellence in Scientific Research, 2015 IEEE Guillemin-Cauer Best Paper Award, 2015 Best Paper of Computer Architecture Letters, Viterbi Fellowship, Jacobs Fellowship, ERC starting grant, the 2017 Pazy Memorial Award, the 2014, 2017 and 2021 Hershel Rich Technion Innovation Awards, 2013 Sanford Kaplan Prize for Creative Management in High Tech, 2010 Benin prize, and seven Technion excellence teaching awards. His current research is focused on circuits and architectures with emerging memory technologies and design of energy efficient architectures.